\documentclass{ws-brl}

\usepackage{graphicx}

\newcommand{\eq}[1]{Eq.~(\ref{#1})}

\newcommand{\bd}{{\rm b}}
\newcommand{\piad}{\pi_{\rm ad}}

\def\siml{\mathrel{\mathchoice {\vcenter{\offinterlineskip\halign{\hfil
$\displaystyle##$\hfil\cr<\cr\sim\cr}}}
{\vcenter{\offinterlineskip\halign{\hfil$\textstyle##$\hfil\cr
<\cr\sim\cr}}}
{\vcenter{\offinterlineskip\halign{\hfil$\scriptstyle##$\hfil\cr
<\cr\sim\cr}}}
{\vcenter{\offinterlineskip\halign{\hfil$\scriptscriptstyle##$\hfil\cr
<\cr\sim\cr}}}}}

\begin{document}

\markboth{Klumpp, M\"uller, and Lipowsky} {Cooperative transport by
  small teams of molecular motors}

\catchline{}{}{}{}{}

\title{COOPERATIVE TRANSPORT BY SMALL TEAMS OF MOLECULAR
  MOTORS}

\author{STEFAN KLUMPP}

\address{Center for Theoretical Biological Physics, University of
  California at San Diego,\\ 9500 Gilman Drive, La Jolla CA
  92109-0374, USA\\
  klumpp@ctbp.ucsd.edu}

\author{MELANIE J. I. M\"ULLER and REINHARD LIPOWSKY}

\address{Max Planck Institute of Colloids and Interfaces, Science Park Golm\\
  14424 Potsdam, Germany 
}

\maketitle

\begin{history}
  \received{25 July 2006} 
\end{history}

\begin{abstract}
Molecular motors power directed transport of cargoes within cells. Even if a single motor 
is sufficient to transport a cargo, motors often cooperate in small teams. We discuss the 
cooperative cargo transport by several motors theoretically and explore some of its 
properties. In particular we emphasize how motor teams can drag cargoes through a 
viscous environment. 
  \keywords{molecular motors; diffusion; active transport; viscous load.}
\end{abstract}

\section{Introduction}

Life is intimately related to movement on many different time and
length scales, from molecular movements to the motility of cells and
organisms. One type of movement which is 
ubiquitous on the molecular
and cellular scale, although not specific to the organic world, is
Brownian motion or passive diffusion: Biomolecules, vesicles,
organelles, and other subcellular particles constantly undergo random
movements due to thermal fluctuations.\cite{Berg1993}  Within cells,
these random movements depend strongly on the size of the diffusing
particles, because the effective viscosity of the cytoplasm increases
with increasing particle size.\cite{Luby-Phelps2000} While proteins
typically diffuse through cytoplasm with diffusion coefficients in the
range of $\mu{\rm m}^2$/s to tens of $\mu{\rm m}^2$/s and therefore explore 
the volume of
a cell  within a few minutes to several tens of minutes (for a typical cell size of 
a few tens of microns), a 100 nm sized organelle typically has a
diffusion coefficient of $\sim 10^{-3}\mu{\rm m}^2$/s within the cell,\cite {Luby-Phelps2000}  
and would need $\sim 10$ days to diffuse over
the length of the cell.

For fast and efficient transport of large cargoes, cells therefore use
active transport based on the movements of molecular motors along
cytoskeletal filaments.\cite{Howard2001,Schliwa2003,Lipowsky_Klumpp2005} 
These molecular
motors convert the chemical free energy released from the hydrolysis
of ATP (adenosinetriphosphate) into directed motion and into
mechanical work. They move in a directed stepwise fashion along the
linear tracks provided by the cytoskeletal filaments. There are three
large families of cytoskeletal motors, kinesins and dyneins which move
along microtubules, and myosins which move along actin filaments. The
filaments have polar structures and encode the direction of motion for
the motors. A specific motor steps predominantly in one direction, the
forward direction of that motor.  Backward steps are usually rare
as long as the motor movement is not opposed by a large force. Motor
velocities are typically of the order of 1 $\mu$m/s, which allows a
motor-driven cargo to move over typical intracellular distances in a
few seconds to a few minutes. 
On the other hand, the force generated by a motor molecule is
of the order of a few pN, which is comparable or larger than estimates for the viscous force
experienced by typical ($\sim$ 100 nm sized) motor-driven cargoes in the cytoplasm.

A large part of our present knowledge about the functioning of
molecular motors is based on {\it in vitro} experiments which have provided
detailed information about the molecular mechanisms of the motors and which have
allowed for systematic measurements of their transport 
properties.\cite{Howard2001} In order to obtain such detailed information, the
overwhelming majority of these experiments has addressed the behavior
of single motor molecules. Within cells, however, transport is often
accomplished by the cooperation of several motors rather than by a
single motor as observed by electron 
microscopy\cite{Miller_Lasek1985,Ashkin__Schliwa1990} and by force measurements\cite{Gross__Wieschaus2002,Gross__Gelfand2002} and the analysis of
cargo particle trajectories {\it in vivo}.\cite{Gross__Wieschaus2002,Gross__Gelfand2002,Hill__Holzwarth2004,Levi__Gelfand2006}
In order to understand the cargo transport in cells, it is therefore
necessary to go beyond the single molecule level and to address how
several motors act together in a team, in particular in cases where
the cooperation of different types of motors is required such as
bidirectional cargo transport. The latter situation, i.e. the presence of
different types of motors bound to one cargo particle, is rather common
and has been observed for kinesins and dyneins, kinesins and myosins
as well as for different members of the kinesin
family and even for members of all three motor families.\cite{Gross2004,Welte2004}

In this article, we review our recent theoretical
analysis\cite{Klumpp_LipowskyCoopTr} of the cooperation of several
motors pulling one cargo. We emphasize the ability of transport driven
by several motors to deal with high viscosities and present an
extended discussion of the case where a strong viscous force opposes
the movement of the cargo particle. We also discuss how diffusion can
be enhanced by motor-driven active transport and conclude with some
remarks on the regulation of active transport.

\section{Stochastic modeling of motor cooperation}

\begin{figure}[tb]
 \begin{center}
    \leavevmode
    \includegraphics[angle=0,width=.9\textwidth]{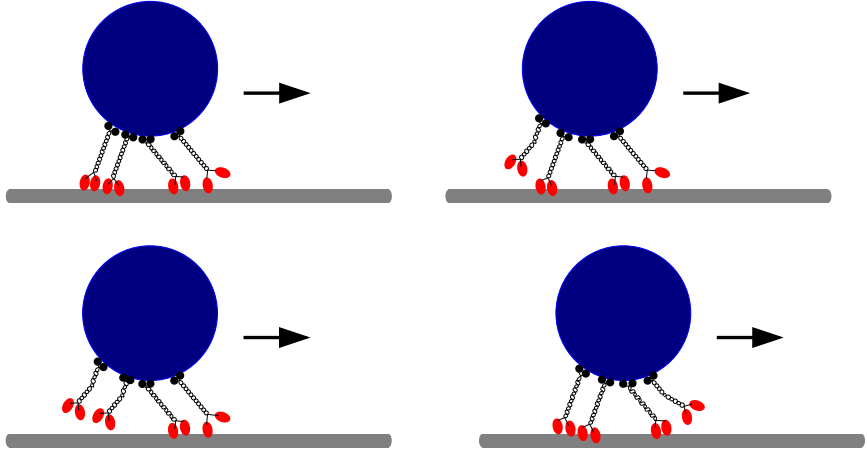}
    \caption{A cargo particle is transported by $N=4$ motors along a cytoskeletal filament. 
    The number of motors which actually pull the cargo changes in a stochastic fashion due 
    to the binding and unbinding of motors to and from the filament.}
    \label{f1}
  \end{center}
\end{figure}

To study the cooperation of several molecular motors theoretically, we
have recently introduced a model which describes the stochastic
binding and unbinding of motors and filaments as well as the movements
of the cargo particle to which these motors are attached.\cite{Klumpp_LipowskyCoopTr} 
The state of the cargo
particle is described by the number $n$ of motors bound to the filament.
As shown in Fig.~\ref{f1}, 
this number changes stochastically between 0 and $N$, the total number
of motors bound to the cargo, since motors bind to and unbind from the
filament.\cite{Lipowsky__Nieuwenhuizen2001} The model is therefore
defined by a set of rates $\epsilon_n$ and $\pi_n$ which describe the
unbinding and binding of a motor, respectively, and which depend on
the number $n$ of bound motors, and by a set of velocities $v_n$ with
which the cargo particle moves when pulled by $n$ motors.

In the simplest case, the motors bind to and unbind from the filament
in a fashion independent of each other. In that case, the binding and
unbinding rates are given by
\begin{equation}
  \label{eq:1}
  \epsilon_n=n\epsilon\qquad {\rm and}\qquad \pi_n=(N-n)\piad
\end{equation}
with the single motor unbinding and binding rates $\epsilon$ and
$\piad$, respectively. For non-interacting motors, the cargo velocity
is independent of the number of pulling motors
and given by the single motor velocity, $v_n=v$, as shown both by
microtubule gliding assays and by bead assays for kinesin motors.\cite{Howard__Vale1989,Coy__Howard1999b,Beeg__LipowskyXXX}  For this
case we have obtained a number of analytical results.\cite{Klumpp_LipowskyCoopTr}
 In particular, the model indicates a
strong increase of the average run length, i.e., the distance a cargo
particle moves along a filament before it unbinds from it. For motors
which bind strongly to the filament, so that $\piad/\epsilon\gg 1$,
the average run length is given by
\begin{equation}
  \label{eq:2}
  \langle\Delta x_\bd\rangle\approx\frac{v}{\epsilon N}(\piad/\epsilon)^{N-1}
\end{equation}
and essentially increases exponentially with increasing number of
motors. Using the single molecule parameters for conventional kinesin
(kinesin 1), we have estimated that run lengths in the centimeter
range are obtained if cargoes are pulled by 7--8 motors.\cite{Klumpp_LipowskyCoopTr} 
As these long
run lengths exceed the length of a microtubule (typically a few tens of
microns), they can however only be realized if microtubules are
aligned in a parallel and isopolar fashion and if cargoes can step
from one microtubule to another as observed {\it in vitro} using aligned
microtubules.\cite{Boehm__Unger2001}  The increase of cargo run
lengths with increasing number of motors has been observed in several
{\it in vitro} experiments,\cite{Coy__Howard1999b,Seitz_Surrey2006,Beeg__LipowskyXXX} 
however it
has been difficult to determine the number of motors pulling the
cargo. One  method to determine the motor number is to use a combination of
dynamic light scattering measurements and comparison of measured run length distributions with theoretical
predictions.\cite{Beeg__LipowskyXXX}

If the cargo is pulled against an opposing force $F$, this force is
shared among the bound motors, so that each bound motor experiences
the force $F/n$. Under the influence of an external force, the single motor velocity decreases
approximately linearly, $v(F)=v(1-F/F_s)$, and the unbinding rate
increases exponentially, $\epsilon(F)=\epsilon \exp(F/F_d)$ as obtained
from optical tweezers
experiments.\cite{Visscher__Block1999,Schnitzer__Block2000} The two
force scales are the stall force $F_s$ and the detachment force $F_d$.
For a cargo pulled by several motors, the velocities and unbinding
rates in the different binding states are then given by
\begin{equation}
  v_n=v\left(1-\frac{F}{nF_s}\right)\qquad{\rm and}\qquad\epsilon_n=n\epsilon\exp\left(\frac{F}{nF_d}\right).
\end{equation}
Since the velocity now depends on the number of bound motors, the
velocity of the cargo changes every time a motor unbinds or an
additional motor binds to the filament. The trajectory of the cargo
therefore consists of linear segments with constant velocity, and the
distribution of the instantaneous velocities has several peaks which
become more and more distinct if the force $F$ is increased. In addition, the
sharing of the force induces a coupling between the motors which leads
to cascades of unbinding events, since the unbinding of one motor
increases the force and, thus, the unbinding rate for the remaining
bound motors. Such unbinding cascades occur also in many other
biophysical systems which have a similar unbinding dynamics, in
particular they have been studied extensively for the forced unbinding
of clusters of adhesion
molecules.\cite{Bell1978,Seifert2000,Erdmann_Schwarz2004}
For the motors, the most important consequence of this type of
force-induced coupling of the motors is that an increase in force not
only slows down the motors, but also decreases the number of bound
motors. Therefore, the force-velocity relation given by the average
velocity as a function of the load force is a nonlinear relation for
cargoes pulled by several motors, although it is approximately linear
for a single motor.\cite{Klumpp_LipowskyCoopTr}

Rather than being imposed by an optical laser trap or other force fields that can be 
directly controlled {\it in vitro}, an opposing force can
also arise from other motors which pull the cargo into the opposite
direction. The presence of two types of motors which move into
opposite directions bound to the same cargo is commonly found in cells
and is required for bidirectional transport in essentially
unidirectional systems of filaments as they are typical for the
microtubule cytoskeleton. In general, the two types of motors interact
both mechanically by pulling on each other \emph{and} via biochemical signals
or regulatory molecules. If there are only mechanical interactions our
model predicts a tug-of-war-like instability: If the motors pull on
each other sufficiently strongly, one species will win, and the cargo
performs fast directed motion rather than being stalled by the pulling
of motors in both directions. Since the number of motors pulling the
cargo is typically small, the direction of motion will however be
reversed from time to time with a reversal frequency which decreases
as the motor numbers are increased.

\section{Motor cooperation in viscous environments}

One universal force that is always experienced by molecular motors 
is the viscous drag caused by the medium through which the cargo is pulled. 
In water or aqueous solutions, however, the viscous drag of the cargo is
usually negligible since it corresponds to a force of only a small
fraction of the motor stall force. For example, a bead with diameter 
1~$\mu$m which moves at 1 $\mu$m/s through water experiences a viscous
force of 0.02 pN which is tiny compared to a motor stall force of a few pN. 
Therefore, {\it in vitro} experiments are hardly affected by the viscosity
of the solution, and changes in motor number do not lead to a change
of the cargo velocity unless the viscosity is increased to $\sim 100$ times that of water.\cite{Hunt__Howard1994}

In highly viscous environments, this is different: If the viscous drag force is of the 
same order of magnitude as the single motor stall force, the velocity can be increased 
if the number of  motors which share this force is increased. 
The latter effect has been observed in microtubule gliding assays with high solution 
viscosity where for low motor density on the surface the velocity decreases as a 
function of the microtubule length, while for high motor density the velocity is 
independent of the microtubule length.\cite{Hunt__Howard1994}

For a cargo pulled by $n$ motors, inserting the Stokes friction force
$F=\gamma v$ (with the friction coefficient $\gamma$) into the
linear force--velocity relation leads to\cite{Klumpp_LipowskyCoopTr}
\begin{equation}\label{eq:vn_gamma}
  v_n(\gamma)=\frac{v}{1+\frac{\gamma v}{n F_s}}\approx \frac{n F_s}{\gamma}. 
\end{equation}
This equation shows that the velocity increases with increasing number of motors 
if $\gamma v/(nF_s)$ is not negligibly small compared to one. In particular, in the limit of 
high viscosity or large
$\gamma v$, for which the last approximation 
in \eq{eq:vn_gamma} is valid, the velocity is proportional to the number $n$ of pulling 
motors.\footnote{Strictly speaking, this proportionality is only valid for small $n$. As a 
consequence of this, an increase of $n$ to motor numbers large compared to 
$\gamma v/F_s$ will increase the consumption of ATP without substantially 
increasing the cargo velocity.} 
In a highly viscous environment, the cargo's velocity distribution therefore exhibits
maxima at integer multiples of a minimal velocity. Similar velocity distributions have 
recently been observed for vesicles and melanosomes in the cytoplasm, see 
Refs.~\refcite{Hill__Holzwarth2004,Levi__Gelfand2006}.\footnote{The microtubule 
gliding assays of Ref.~\refcite{Hunt__Howard1994} with high viscosity and intermediate 
motor densities on the surface exhibit a large variability of the velocity, however discrete 
peaks have not  been resolved in that experiment.} 
To first order in $\gamma^{-1}$ the motors experience the
force $F_n\approx n F_s$ which implies that the force per bound motor
is independent of the number of bound motors and that the motors
behave as independent motors for large viscous force, however with an
increased effective single motor unbinding rate $\epsilon \exp(F_s/F_d)$.

\section{Active diffusion: Motor-driven diffusive movements}

We have emphasized that large particles experience a strong viscous
drag in the cytoplasm and that therefore Brownian motion is too slow
to drive transport of large particles in the cell. While this
observation suggests that active transport is necessary within cells,
it does not imply that the active transport must necessarily be
directed transport. Alternatively, active transport could also be used
to generate effectively diffusive motion, which is faster than passive
Brownian motion, e.g. if a cargo particle performs a sequence of
active molecular motor-driven runs in random direction. We call this
effectively diffusive motion, which depends on chemical energy, active
diffusion.\cite{Klumpp_LipowskyAct} In cells, active diffusion can be
achieved either by (i) switching
the direction of motion by switching between different types of motors which walk 
along a unipolar array of filaments
or by (ii) a single type of motor and isotropic (e.g., bidirectional
or random) arrangements of filaments. The first case is typical for
microtubule-based transport: microtubules are often arranged in a
directed fashion, either in radial systems emanating from a central
microtubule organizing center with their plus ends pointing outwards
or in unidirectional systems where microtubules are aligned in a
parallel and isopolar fashion such as in axons.\footnote{For these two types of 
filament alignments, we have recently determined the stationary motor concentration 
profiles, see Ref.~\refcite{Klumpp__Lipowsky2005}.} Bidirectional
movements along these unidirectional microtubule arrangements have been 
observed for a large variety of intracellular cargoes.\cite{Gross2004,Welte2004} 
These movements are driven by a combination of plus end and
minus end directed motors.  On the other hand, actin-based movements
are often of the second type, since the actin cytoskeleton usually forms an isotropic 
random mesh, on which, e.g., myosin V-driven cargoes
perform random walks.\cite{Snider__Gross2004,Dinh__Mitragotri2006}
Active diffusion has also been observed for random arrays of microtubules in cell 
extracts.\cite{Salman__Elbaum2005}  Let us mention that these two types of
active diffusion are highly simplified. More complex scenarios include
bidirectional, but biased movements along microtubules and the switching of 
cargoes between
microtubules and actin filaments.\cite{Gross2004,Welte2004}  In
vitro, one can use various techniques such as chemically or
topographically structured surfaces,\cite{Hess_Vogel2001} 
motor--filament self-organization,\cite{Surrey__Karsenti2001} and
filament crosslinking on micropillars\cite{Roos__Spatz2003} to create
well-defined patterns of filaments for active diffusion which may be
useful to enhance diffusion in bio-nanotechnological transport 
systems.\cite{Klumpp_LipowskyAct}

The maximal effective diffusion coefficient which can be achieved by
molecular motor-driven active diffusion is given by
\begin{equation}
  D_{\rm act}\approx v L P_\bd.
\end{equation}
where $L$ is the length of essentially unidirectional runs, given by
either the average run length before unbinding from filaments or the
mesh size of the filament pattern, and $P_\bd$ is the probability that
the cargo is bound to a filament.\cite{Klumpp_LipowskyAct} In order to obtain 
large effective
diffusion one therefore has to make sure that the cargo particle is
bound to filaments most of the time, and that it has an average run
length which is comparable or larger than the pattern mesh size. One
possibility to satisfy both conditions is to use a sufficiently large
number of motors. A larger number of motors also decreases the
probability of switching direction at an intersection,\cite{Snider__Gross2004} so that 
unidirectional runs exceeding the mesh size can be achieved.

Since the motor velocity is rather insensitive to the viscosity for
small viscosities, active diffusion is much less affected by the
viscous drag of the solution than passive Brownian motion,\cite{Klumpp_LipowskyAct} 
in particular if a cargo is pulled by
several motors. For a cargo of size 100 nm, Brownian motion in water
is characterized by a diffusion coefficient in the order of a few
$\mu$m$^2$/s. An increase in viscosity by a factor of 10 leads to a
decrease of the diffusion coefficient by that factor. The active
diffusion coefficient, on the other hand can be estimated to be of
similar size or slightly smaller (using $v\sim 1\mu$m/s, $L\sim 1$--10~$\mu$m
and $P_\bd\siml 1$), however the latter value is essentially
unaffected by an increase of the viscosity by up to a factor 100
compared to the viscosity of water, since the viscous force which
arises from the movement of such a bead is only a fraction of $\sim
10^{-3}$ of the motor stall force. The effect is increased if a cargo
is pulled by several motors, since the viscous force $\gamma v$ starts
to affect the motor movement only if it is of the order $\langle n\rangle
F_s$. Therefore only viscosities which lead to viscous forces that
exceed $\langle n\rangle F_s$ have a considerable effect on active
diffusion.

\section{Aspects of control}

In the preceding sections we have discussed how cells achieve cargo
transport over large distances through a viscous environment by the
cooperation of a small number of molecular motors. In principle, cells
could also use a single motor which generates a larger force and binds
more strongly to the corresponding filament rather than a team of
motors, but motor cooperation appears to be preferred. The use of
several motors has the advantage that the transport parameters can be
easily controlled by controlling the number of pulling motors, e.g. by
activating motors, by increasing their binding to filaments or by
recruiting additional motors to the cargo.

The use of multiple weak bonds rather than a single strong
bond in order to enable simple ways of control appears to be a general
principle in cellular biology which applies to various cellular
processes as diverse as the build-up of strong, but at the same time
highly dynamic clusters of adhesion molecules\cite{Schwarz__Bischofs2006} 
and the binding of transcription factors
to DNA where programmable specificity of binding is achieved by
sequence-dependent binding of the transcription factor to a short
stretch of DNA.\cite{Gerland__Hwa2002}

If it is true that the main purpose of motor cooperation is the
controllability of motor-driven movements, one may ask whether this
function imposes constraints on the properties of the motors. 
For example, in order to both up- and down-regulate the cargo velocity or run length,         
it is clearly desirable to be able to both increase and decrease the            
number                                                                          
of pulling motors and, thus, to have an average number, $\langle n \rangle$,                                                                       
of pulling motors that is not too close to either 1 or $N$,
the total number of motors, but rather is of the order of $N/2$.\footnote{As mentioned, in 
highly viscous environments the total motor number $N$ itself may be fixed by the 
tradeoff between increased cargo velocity and efficient ATP usage, such that 
$\langle n\rangle\sim \gamma v/F_s$.} The
average number of pulling motors can be estimated\cite{Klumpp_LipowskyCoopTr} 
by $\langle n\rangle\sim N/[1+\epsilon(F)/\piad]$, so that the requirement 
$\langle n\rangle \sim N/2$ implies $\epsilon(F)/\piad\sim 1$ or, 
 for cargoes subject to a strong
viscous force, $\piad/\epsilon\sim \exp(F_s/F_d)$. The latter condition represents a 
relation between the
single motor parameters possibly imposed by a functional constraint
related to motor cooperation. The parameters of conventional kinesin
approximately satisfy this relation, which could however be
coincidental. It would therefore be interesting to see whether this relation is
also satisfied by the parameters of other motors. If this is not the
case, these differences might provide hints towards functional
differences between different types of motors.

\section{Concluding remarks}

Cargo transport in cells is often carried out by small teams of molecular motors 
rather than by single motor molecules. We have described how one can analyze the 
transport by several motors using a phenomenological model based on the 
understanding of single motors that has been obtained from single molecule 
experiments during the last decade. Our theoretical approach is sufficiently versatile 
to be extended to more complex situations where additional molecular species are 
present such as the cooperation of two types of motors or the interaction of motors 
with regulatory proteins. 
In particular, it will be interesting to use this model to study control mechanisms of 
motor-driven transport.

\section*{Acknowledgments}

SK acknowledges support by Deutsche Forschungsgemeinschaft (KL 818/1-1).


\begin{thebibliography}{10}

\bibitem{Berg1993}
H.~C. Berg,
\newblock {\em Random walks in biology},
\newblock 2nd end. (Princeton University Press, Princeton/Chichester, 1993).

\bibitem{Luby-Phelps2000}
K.~Luby-Phelps,
\newblock {\em Int.\ Rev.\ Cytol. -- Survey Cell Biol.}  {\bf 192}, 189 (2000).

\bibitem{Howard2001}
J.~Howard,
\newblock {\em Mechanics of Motor Proteins and the Cytoskeleton},
\newblock (Sinauer Associates, Sunderland, 2001).

\bibitem{Schliwa2003}
M.~{Schliwa} (ed.),
\newblock {\em Molecular motors}
\newblock (Wiley-VCH, Weinheim, 2003).

\bibitem{Lipowsky_Klumpp2005}
R.~Lipowsky and S.~Klumpp,
\newblock {\em Physica A} {\bf 352}, 53 (2005).

\bibitem{Miller_Lasek1985}
R.~H. Miller and R.~J. Lasek,
\newblock {\em J.\ Cell Biol.} {\bf 101}, 2181 (1985).

\bibitem{Ashkin__Schliwa1990}
A.~Ashkin, K.~Sch{\"u}tze, J.~M. Dziedzic, U.~Euteneuer, and M.~Schliwa,
\newblock {\em Nature} {\bf 348}, 346 (1990).

\bibitem{Gross__Wieschaus2002}
S.~P. Gross, M.~A. Welte, S.~M. Block, and E.~F. Wieschaus,
\newblock {\em J. Cell Biol.}  {\bf 156}, 715 (2002).

\bibitem{Gross__Gelfand2002}
S.~P. Gross, M.~C. Tuma, S.~W. Deacon, A.~S. Serpinskaya, A.~R. Reilein, and
  V.~I. Gelfand,
\newblock {\em J. Cell Biol.} {\bf 156}, 855 (2002).

\bibitem{Hill__Holzwarth2004}
D.~B. Hill, M.~J. Plaza, K.~Bonin, and G.~Holzwarth,
\newblock {\em Eur.\ Biophys.\ J} {\bf 33}, 623 (2004).

\bibitem{Levi__Gelfand2006}
V.~Levi, A.~S. Serpinskaya, E.~Gratton, and V.~Gelfand,
\newblock {\em Biophys. J.} {\bf 90}, 318 (2006).

\bibitem{Gross2004}
S.~P. Gross,
\newblock {\em Phys. Biol.} {\bf 1}, R1 (2004).

\bibitem{Welte2004}
M.~A. Welte,
\newblock {\em Curr.\ Biol.} {\bf 14}, R525 (2004).

\bibitem{Klumpp_LipowskyCoopTr}
S.~Klumpp and R.~Lipowsky,
\newblock {\em Proc.\ Natl.\ Acad.\ Sci.\ USA} {\bf 102}, 17284 (2005).

\bibitem{Lipowsky__Nieuwenhuizen2001}
R.~Lipowsky, S.~Klumpp, and Th.~M. Nieuwenhuizen,
\newblock {\em Phys.\ Rev.\ Lett.} {\bf 87}, 108101 (2001).

\bibitem{Howard__Vale1989}
J.~Howard, A.~J. Hudspeth, and R.~D. Vale,
\newblock {\em Nature} {\bf 342}, 154 (1989).

\bibitem{Coy__Howard1999b}
D.~L. Coy, M.~Wagenbach, and J.~Howard,
\newblock {\em J.\ Biol.\ Chem.} {\bf 274}, 3667 (1999).

\bibitem{Beeg__LipowskyXXX}
J.~Beeg et~al.,
\newblock Manuscript in preparation.

\bibitem{Boehm__Unger2001}
K.~J. B{\"o}hm, R.~Stracke, P.~M{\"u}hlig, and E.~Unger,
\newblock {\em Nanotechnology} {\bf 12}, 238 (2001).

\bibitem{Seitz_Surrey2006}
A.~Seitz and T.~Surrey,
\newblock {\em EMBO J.} {\bf 25}, 267 (2006).

\bibitem{Visscher__Block1999}
K.~Visscher, M.~J. Schnitzer, and S.~M. Block,
\newblock {\em Nature} {\bf 400}, 184 (1999).

\bibitem{Schnitzer__Block2000}
M.~J. Schnitzer, K.~Visscher, and S.~M. Block,
\newblock {\em Nature Cell Biol.} {\bf 2}, 718 (2000).

\bibitem{Bell1978}
G.~I. Bell,
\newblock {\em Science} {\bf 200}, 618 (1978).

\bibitem{Seifert2000}
U.~Seifert,
\newblock {\em Phys.\ Rev.\ Lett.} {\bf 82}, 2750 (2000).

\bibitem{Erdmann_Schwarz2004}
T.~Erdmann and U.~S. Schwarz,
\newblock {\em Phys.\ Rev.\ Lett.} {\bf 92}, 108102 (2004).

\bibitem{Hunt__Howard1994}
A.~J. Hunt, F.~Gittes, and J.~Howard,
\newblock {\em Biophys.\ J.} {\bf 67}, 766 (1994).

\bibitem{Klumpp_LipowskyAct}
S.~Klumpp and R.~Lipowsky,
\newblock {\em Phys.\ Rev.\ Lett.} {\bf 95}, 268102 (2005).

\bibitem{Klumpp__Lipowsky2005}
S.~Klumpp, Th.~M. Nieuwenhuizen, and R.~Lipowsky,
\newblock {\em Biophys.\ J.} {\bf 88}, 3118 (2005).

\bibitem{Snider__Gross2004}
J.~Snider, F.~Lin, N.~Zahedi, V.~Rodionov, C.~C. Yu, and S.~P. Gross,
\newblock {\em Proc.\ Natl.\ Acad.\ Sci.\ USA} {\bf 101}, 13204 (2004).

\bibitem{Dinh__Mitragotri2006}
A.-T. Dinh, C.~Pangarkar, T.~Theofanous, and S.~Mitragotri,
\newblock {\em Biophys. J.} {\bf 90}, L67 (2006).

\bibitem{Salman__Elbaum2005}
H.~Salman, A.~Abu-Arish, S.~Oliel, A.~Loyter, J.~Klafter, R.~Granek, and
  M.~Elbaum,
\newblock {\em Biophys. J.} {\bf 89}, 2134 (2005).

\bibitem{Hess_Vogel2001}
H.~Hess and V.~Vogel,
\newblock {\em Rev.\ Mol.\ Biotechnology} {\bf 82}, 67 (2001).

\bibitem{Surrey__Karsenti2001}
T.~Surrey, F.~{N\'ed\'elec}, S.~Leibler, and E.~Karsenti,
\newblock {\em Science} {\bf 292}, 1167 (2001).

\bibitem{Roos__Spatz2003}
W.~H. Roos, A.~Roth, J.~Konle, H.~Presting, E.~Sackmann, and J.~P. Spatz,
\newblock {\em Chem.\ Phys.\ Chem.} {\bf 4}, 872 (2003).

\bibitem{Schwarz__Bischofs2006}
U.~S. Schwarz, T.~Erdmann, and I.~B. Bischofs,
\newblock {\em BioSystems} {\bf 83}, 225 (2006).

\bibitem{Gerland__Hwa2002}
U.~Gerland, J.~D. Moroz, and T.~Hwa,
\newblock {\em Proc.\ Natl.\ Acad.\ Sci.\ USA} {\bf 99}, 12015 (2002).

\end{thebibliography}
\end{document}